\documentclass[11pt]{article}
\usepackage{hyperref}
\usepackage{amsmath,amssymb,amsthm,amscd}
\usepackage[all,cmtip]{xy}
\usepackage{xcolor}
\usepackage{blkarray}

\numberwithin{equation}{section}

\begin{document}
\begin{titlepage}
{}~ \hfill\vbox{ \hbox{} }\break

\rightline{UWThPh-2019-06}

\vskip 3.5 cm

\centerline{\Large \bf Modularity from Monodromy}  \vskip 0.5 cm
\centerline{\Large \bf  }   \vskip 0.5 cm

\renewcommand{\thefootnote}{\fnsymbol{footnote}}
\vskip 20pt \centerline{ 
Thorsten Schimannek\footnote{thorsten.schimannek@univie.ac.at} } \vskip .5cm \vskip 20pt

\begin{center}

\textit{Faculty of Physics, University of Vienna}\\\textit{Boltzmanngasse 5, A-1090 Vienna, Austria}\\ [3 mm]

\end{center}

\setcounter{footnote}{0}
\renewcommand{\thefootnote}{\arabic{footnote}}
\vskip 60pt
\begin{abstract}
	In this note we describe a method to calculate the action of a particular Fourier-Mukai transformation on a basis of brane charges on elliptically fibered Calabi-Yau threefolds with and without a section.
	The Fourier-Mukai kernel is the ideal sheaf of the relative diagonal and for fibrations that admit a section this is essentially the Poincaré sheaf.
	We find that in this case it induces an action of the modular group on the charges of 2-branes.
\end{abstract}

\end{titlepage}
\vfill \eject

\newpage

\baselineskip=16pt    

\tableofcontents
\section{Introduction}
F-theory on elliptically fibered Calabi-Yau varieties is one of the richest sources of semi-realistic string compactifications~\cite{Vafa:1996xn,Morrison:1996na,Morrison:1996pp}.
This is facilitated by the fact that the vast majority of known Calabi-Yau threefolds seem to be elliptically fibered~\cite{Gray:2014fla,Anderson:2017aux,Huang:2018esr} and that
elliptically fibered Calabi-Yau $n$-folds with desirable properties can be engineered using tools from toric geometry~\cite{Cvetic:2013nia,Klevers:2014bqa}.
In this way F-theory leads to a rich dictionary between the geometric and arithmetic structure of elliptic fibrations on one side and properties of effective theories of quantum gravity on the other.

In particular, the dictionary relates the topological string partition function which encodes enumerative invariants of the Calabi-Yau to the elliptic genera of certain strings.
The latter arise from D3-branes that wrap curves in the base of the fibration.
From the transformation of elliptic genera under Dehn twists of the worldsheet one can then deduce that the topological string partition function should exhibit certain modular properties~\cite{Schellekens:1986xh,Schellekens:169145,Klemm:1996hh,Haghighat:2013gba}.
The details of the relation between geometric properties of the Calabi-Yau and modular properties of the elliptic genera have recently been used to provide highly non-trivial evidence for a set of ``swampland conjecture'' ~\cite{Lee:2018urn,Lee:2018spm}.
The swampland program is an ongoing effort to understand the sine qua non of quantum field theories that can consistently be embedded into theories of quantum gravity.

However, modular properties of the genus zero contribution to the topological string partition function on elliptically fibered Calabi-Yau threefolds have already been observed in~\cite{Candelas:1994hw}.
There it arose in the context of mirror symmetry.

Mirror symmetry relates the periods over the holomorphic $(3,0)$-form on a Calabi-Yau $W$ to the leading part of the A-model topological string partition function on the mirror $M$.
Moreover, a subset of the periods can be used as flat coordinates on the complex structure moduli space of $W$ that provide a local identification with the (quantum) K\"ahler moduli space of $M$.
The authors of~\cite{Candelas:1994hw} determined the full complex structure monodromy group for the mirror of $X_{18}$, a degree $18$ hypersurface in the weighted projective space $\mathbb{P}(1,1,1,6,9)$.
Two monodromies $T,S$ act on the flat coordinate $\tau$ that is under mirror symmetry identified with the complexified volume of the fiber of $X_{18}$ via
\begin{align}
	T:\,\tau\mapsto\tau+1\,,\quad S:\,\tau\mapsto-\frac{1}{\tau}\,.
	\label{eqn:x18trans1}
\end{align}
An appropriate combination of flat coordinates $t=\tilde{t}+\frac32\tau$ where $t$ is identified with the volume of a curve in the base of the fibration transforms in the limit of large base as
\begin{align}
	T:\,t\mapsto t+\frac{3}{2}\,,\quad S:\,t\mapsto t-\frac{3}{2}\,.
	\label{eqn:x18trans2}
\end{align}
From this action they were able to deduce that the Yukawa-coupling $C_{\tau\tau\tau}$ is in the limit $\text{Im}t\rightarrow \infty$ proportional to the Eisenstein series $E_4(\tau)$.

In subsequent works the understanding of the modular properties of the topological string partition function on non-compact~\cite{Hosono:1999qc,Hosono:2002xj} and compact elliptic fibrations has been significantly improved upon~\cite{Klemm:2012sx,Alim:2012ss}.
Most importantly it was found that the modular structure extends to arbitary base degree and that the holomorphic anomaly equations found by~\cite{Bershadsky:1993cx} imply a ``modular recursion'' in terms of the base degree and the worldsheet genus.
A culmination of this effort was the HKK-conjecture which states that the topological string partition function on $X_{18}$ can be expanded in particular quotients of weak Jacobi forms~\cite{Huang:2015sta} (see also \cite{Oberdieck:2016nvt,Oberdieck:2017zit,Oberdieck:2017pqm} for a mathematical discussion and a partial proof).
In particular it was argued that the appearance of weak Jacobi forms is a consequence of the action \eqref{eqn:x18trans1}, \eqref{eqn:x18trans2} and the modular anomaly equation~\cite{Huang:2015sta,Katz:2016SM}.
Exploiting this structure it was possible to obtain the first all-genus results for compact Calabi-Yau varieties~\cite{Huang:2015sta}.

Until recently the study of modular properties of the topological string partition function on elliptic Calabi-Yau has mostly been restricted to geometries that have only one section and where the rest of the divisors
arise form pull-backs of curves in the base.
Further generalizations addressed geometries with divisors that resolve curves of ADE-singularities~\cite{DelZotto:2017mee} and with multiple sections~\cite{Lee:2018urn,Lee:2018spm}.
However, these relied exclusively on the intricate duality with elliptic genera of strings and did not connect to the monodromies of the Calabi-Yau.
Part of the reason is perhaps that a detailed study of the monodromies of multi-moduli families of Calabi-Yau varieties is rather cumbersome.

A shortcut is provided by the homological mirror symmetry conjecture~\cite{Kontsevich:1994dn} which states that mirror symmetry identifies the category of B-branes on a Calabi-Yau $M$ with the category of A-branes on the mirror $W$.
In particular the central charges of a mirror pair of branes should be identified via the mirror map.
A-branes correspond to special lagrangian cycles in $W$ and the central charge of an A-brane is given by the integral of the holomorphic $(3,0)$-form over the cycle that supports the brane.
On the other hand, B-branes on $M$ are objects in the bounded derived category of quasi-coherent sheaves $D^b(M)$.
Monodromies in the complex structure moduli space of $W$ act on the A-branes on $W$ and thus, via mirror symmetry, on the B-branes on $M$.
The homological mirror symmetry conjecture also proposes that the action of the monodromies lift to an autoequivalence of $D^b(M)$.

A theorem by Orlov~\cite{Orlov:1996} states that equivalences of derived categories $D^b(X),D^b(Y)$ are always expressible as Fourier-Mukai transformations
\begin{align}
	\Phi_{\mathcal{E}}:\,\mathcal{F}^\bullet\mapsto R\pi_{1*}\left(\mathcal{E}\otimes_L L\pi_2^*\mathcal{F}^\bullet\right)\,,
	\label{eqn:fmtransform}
\end{align}
where $\pi_i,i=1,2$ are the projections from $Y\times X$ to the $i$-th factor and the so-called Fourier-Mukai kernel $\mathcal{E}$ is a sheaf on $Y\times X$. The letters $L$ and $R$ indicate that the corresponding derived version of a functor is to be taken.
Autoequivalences correspond to the special case $X=Y$.
Using the Grothendieck-Riemann-Roch theorem one can relate the transformation \eqref{eqn:fmtransform} to an action on the brane charges in cohomology~\cite{huybrechts2006fourier}.
For several important B-model monodromies the corresponding Fourier-Mukai kernel is known to be of a generic form which in principal allows one to calculate the corresponding action on the branes.
Most prominently the kernel that corresponds to a monodromy around the generic point of the principal component of the discriminant is conjectured to be the ideal sheaf of the diagonal~\cite{Horja:1999,Morrison:2000bt}.

In fact, the $T$ transformation in \eqref{eqn:x18trans1},\eqref{eqn:x18trans2} is a large volume/large complex structure monodromy and the kernel is $j_*\mathcal{O}(\tau)$.
The result of the corresponding action \eqref{eqn:fmtransform} on a complex of sheaves $\mathcal{F}\in D^b(X_{18})$ is just $\mathcal{F}\otimes\mathcal{O}(\tau)$.
On the other hand, the autoequivalence of $D^b(X_{18})$ that leads to the $S$ transformation is significantly more interesting.
It can be written as a \textit{relative} Fourier-Mukai transformation where the projections in \eqref{eqn:fmtransform} refer to the relative fiber product $X\times_B X$.
The corresponding kernel is the Poincaré sheaf~\cite{Andreas:2000sj,Andreas:2004uf}
\begin{align}
	\mathcal{P}=\mathcal{I}_\Delta\otimes\pi_1^*\mathcal{O}_{X_{18}}(\tau)\otimes\pi_2^*\mathcal{O}_{X_{18}}(\tau)\otimes q^*c_1(B)\,,
	\label{eqn:x18poincaresheaf}
\end{align}
where $q$ is the projection from $X\times_B X$ to $B$ and $\mathcal{I}_\Delta$ is the ideal sheaf of the relative diagonal.

The transformation induced by the Poincaré sheaf can intuitively be interpreted as T-dualizing both cycles of the fiber.
It is also the transformation that underlies the spectral cover construction introduced in~\cite{Friedman:1997yq}.
The action on elliptic fibrations with one section and without reducible fibers has been discussed on several occasions in the literature, see e.g. the recent paper~\cite{Corvilain:2018lgw}.
However, we are not aware of any explicit calculation in the context of elliptic fibrations with multiple sections, additional fibral divisors or only multi-sections.
This applies in particular to the induced action on the flat coordinates.

Before we make an attempt to remedy this situation let us first replace the kernel \eqref{eqn:x18poincaresheaf} with an object that is also available for fibrations without a section.
To this end we note that the tensor product with line bundles is essentially a choice of normalization and can be compensated by large volume monodromies.
The non-trivial part of the transformation is therefore induced by the ideal sheaf $\mathcal{I}_\Delta$ of the \textit{relative} diagonal.
Indeed it was proven that $\mathcal{I}_\Delta$ induces an auto-equivalence of the derived category for arbitrary genus one fibrations~\cite{ruiperez2006relative}.
We will therefore focus on calculating the action that is induced by $\mathcal{I}_\Delta$ on the brane charges.
In analogy with the ordinary conifold monodromy we will call this a \textit{relative conifold transformation}.

One technical hurdle that has to be overcome is due to singularities.
While we will always assume that the Calabi-Yau itself is smooth this does not hold for all fibers.
A generalization of the Grothendieck-Riemann-Roch theorem to singular varieties has been obtained by Fulton~\cite{fulton1984intersection}.
It was used in~\cite{Andreas:2000sj,Andreas:2004uf} to obtain the action that is induced by the Poincaré sheaf on a basis of brane charges on a general elliptic Calabi-Yau $X$ without reducible fibers.
However, the calculation relies on the isomorphism between $X$ and the corresponding Weierstrass fibration.
This is because the singular Grothendieck-Riemann-Roch theorem requires the fibration $X$ to be a local complete intersection morphism (l.c.i.) in the sense of~\cite{fulton1984intersection}.
For a smooth Weierstrass fibration $\pi:X\rightarrow B$ can be factored into an inclusion into a $\mathbb{P}(1,2,3)$ fibration over $B$ followed by the corresponding projection onto $B$ which makes $\pi$ an l.c.i. morphism.

We will overcome this problem by constructing the fibrations from fibers that are hypersurfaces in more general toric ambient spaces.
The price will be that for different structures of the fibration we have to perform the calculation using a different ambient space.
Our strategy is therefore to explain the calculation at the hand of two examples and highlight the aspects that are generic.
The calculation can then easily be adapted to other families of fibers and even to complete intersections in higher dimensional toric ambient spaces.
The vast amount of fibration structures that can be constructed already from toric hypersurfaces~\cite{Klevers:2014bqa} and complete intersections in three dimensional toric ambient spaces~\cite{Braun:2014qka,Oehlmann:2016wsb} makes us confident
that we have little loss of generality.

After discussing the examples we will calculate the action of the transformation on the 2-brane charges for arbitrary elliptically fibered Calabi-Yau varieties that can be constructed in the previously described manner.
This leads to a generalization of the transformations \eqref{eqn:x18trans1} and \eqref{eqn:x18trans2} (see equation \eqref{eqn:modtrans1}).
If the geometry admits a section we find that the volume of the fiber transforms like a modular parameter while the volumes of rational fibral curves transform like elliptic parameters of a Jacobi form.
The exponential of appropriately shifted base parameters transforms, up to a multiplier system, like a lattice Jacobi form of weight $0$.
The index matrix is block diagonal such that the volumes of isolated fibral curves and fibers of fibral divisors do not mix.
The indices of the former correspond to the height pairing between the sections while the indices of the latter are essentially given by the intersection of the corresponding fibral divisors.

In a companion paper that is currently being finalized we will also study the modular anomaly equation~\cite{wip}.
Combined with the transformations that are obtained in this note we will argue that the coefficients in an expansion of the topological string partition function with respect to appropriately shifted base parameters
are weak Jacobi forms of weight $0$. 
The index matrix with respect to the volumes of fibral curves will be determined by the index matrix of the corresponding power of exponentiated base parameters.
This confirms and generalizes a conjecture that has been made by~\cite{Lee:2018urn,Lee:2018spm} about the properties of the topological string partition function on elliptic fibrations with multiple sections.
In \cite{wip} we will also discuss the action on the 2-brane charges for elliptic fibrations without a section.

The title of this note still needs justification.
While it is conjectured that every monodromy lifts to an autoequivalence of the derived category the converse does not hold in general.
In the case of the quintic the shift functor $\mathcal{F}\mapsto\mathcal{F}[1]$ is not the lift of a monodromy, only the shift-by-two functor is~\cite{Aspinwall:2001dz}.
However, for $X_{18}$ it is known that the relative conifold transformation arises from the \textit{wall monodromy} around the boundary between the geometric phase and a hybrid phase where the fiber collapses~\cite{Horja:1999,Aspinwall:2001zq}.
We verified this fact for $X_{18}$ and an elliptic fibration without a section using matrix factorization techniques in gauged linear sigma models~\cite{Erkinger:2017aaa}.
Using the relation between the wall monodromy, the large volume monodromies and the monodromy around a generic point on the principal component of the discriminant~\cite{Aspinwall:2001zq} we also
checked this for several geometries with reducible fibers.
We conjecture that this relation is general and that the relative conifold transformation \textit{always} arises from a wall monodromy around the boundary between the geometric phase and a hybrid phase where the fiber collapses.

Let us finally note that we use the term elliptic fibration without implying the existence of a section.
We will occasionally use the F-theory dictionary to interpret geometrical properties in terms of the effective quantum field theory.
A recent review is~\cite{Weigand:2018rez}.
Moreover, we will adopt the following nomenclature that is frequently used in the F-theory literature:
A \textit{fibral divisor} is a rational fibration that resolves a singularity in the fiber. A \textit{fibral curve} is a rational curve with support over a point in the base. \textit{Vertical divisors} are pull-backs of divisors from the base.
\section{Calculating the relative conifold transformation}
In this section we will give a detailed explanation of how the action of the relative conifold transformation on the brane charges can be calculated.
We think that it is best to do this at the hand of two examples.
To this end we use the techniques from~\cite{Klevers:2014bqa} and construct elliptic fibrations with $I_2$ fibers over arbitary bases $B$.
The toric construction of the fiber will automatically lead to a toric resolution of the singularity.
Over a curve in the base the fibers of the resulting fibrations are reducible and consist of two rational curves that intersect like the affine Dynkin diagram of $SU(2)$.
Our second example will also exhibit isolated fibral curves.

For a short introduction to toric geometry we recommend~\cite{Cox:2000vi}.
A review on B-branes can be found in~\cite{Aspinwall:2004jr}.
Note however that the discussion of central charges in~\cite{Aspinwall:2004jr} predates the $\Gamma$-class formula~\cite{Iritani:2009} that we use in section \ref{sec:3}).

\subsection{$G=SU(2)/\mathbb{Z}_2\,,\quad \#n_F=0$}
As our first example we construct families of elliptic Calabi-Yau threefolds with a curve of $SU(2)$ singularities and without additional hypermultiplets in the fundamental representation that would arise from isolated fibral curves.
To this end we consider a generic hypersurface in the partially resolved toric ambient space $\mathbb{P}_{F_{13}}$ with the data
\begin{align}
\begin{blockarray}{crrrrl}
	&&&C_F^1&C_F^2&\\
\begin{block}{r(rr|rr)l}
	e_5& -1&-2& 1& 0&\leftarrow\text{ zero-section }E_1\\
	e_4& -1& 2&-1& 1&\leftarrow\text{ torsional section }E_2\\
	  v&  0& 1& 4&-2&\leftarrow\text{ fibral divisor }D_f\\
	  w&  1& 0& 0& 1&\leftarrow\text{ two-section}\\
	   &  0& 0&-4& 0&\\
\end{block}
\end{blockarray}\,.
	\label{eqn:toricdataf13}
\end{align}
On the left we provide the generators of the rays of the toric fan. On the right are the corresponding intersections with a set of curves $C_F^1,C_F^2$ that generate the Mori cone on $\mathbb{P}_{F_{13}}$.
The Stanley-Reisner ideal is
\begin{align}
	\mathcal{SRI}=\langle ve_5,\,we_4\rangle\,.
\end{align}
To obtain an elliptic fibration $M$ we promote the homogeneous coordinates $e_5,e_4,v,w$ to sections of line bundles over a base $B$ and consider vanishing loci of sections of the anti-canonical bundle in the
total space $\widehat{\mathbb{P}}_{F_{13}}\rightarrow B$.
We assume that $B$ is a surface that does not introduce additional singularities from curves with self-intersection greater than two.
We will also assume that the Mori cone of $B$ is simplicial.
The techniques apply equally well to the non-simplicial case but this assumption allows us to write down generic formulas.

Roughly following the conventions of \cite{Klevers:2014bqa} we denote the corresponding divisor classes by
\begin{align}
	\begin{split}
	[v]=&2E_1-2E_2+c_1(B)-\mathcal{S}_7\,,\quad [w]=E_1+E_2+c_1(B)-\mathcal{S}_9\,,\\
	[e_4]=&E_2\,,\quad [e_5]=E_1\,.
	\end{split}
\end{align}
Here we have used the torus action to make $e_4$ and $e_5$ sections of the trivial bundle and  $\mathcal{S}_7,\mathcal{S}_9$ are the first Chern classes of line bundles on $B$.
We also introduced $c_1(B)$ to denote the pull-back of the first Chern class of the tangent bundle of $B$ to $M$.
A generic section of the anti-canonical bundle on $\mathbb{P}_{F_{13}}$ is of the form
\begin{align}
	p_{F_{13}}=s_1e_5^4+s_2e_4^2e_5^2v+s_3e_4^4v^2+s_6e_4e_5vw+s_9vw^2\,,
	\label{eqn:p1}
\end{align}
and for $M$ to be Calabi-Yau we have to promote the coefficients $s_i,\,i\in\{1,2,3,6,9\}$ to sections of the bundles
\begin{align}
	\begin{split}
	[s_1]=&3c_1(B)-\mathcal{S}_7-\mathcal{S}_9\,,\quad [s_2]=2c_1(B)-\mathcal{S}_9\,,\quad [s_3]=c_1(B)+\mathcal{S}_7-\mathcal{S}_9\,,\\
	[s_6]=&c_1(B)\,,\quad [s_9]=\mathcal{S}_9\,.
	\end{split}
\end{align}
Degenerations of the fiber over loci of codimension one and two in $B$ have been classified in \cite{Klevers:2014bqa}.
Using that classification it is easy to check that for $\mathcal{S}_7=-c_1(B),\mathcal{S}_9=0$ the Calabi-Yau $M$ will be smooth.

There is a genus $g=1+6c_1(B)^2$ curve $C_G$ of $I_2$ fibers that are resolved by the fibral divisor $D_f$.
We will denote the two components of a generic fiber over $C_G$ by $C_A$ and $C_B$.
The fibration also exhibits two holomorphic sections $\{e_5=0\}$ and $\{e_4=0\}$ that respectively intersect $C_A$ and $C_B$.
We choose $\{e_5=0\}$ with $[e_5]=E_1$ to be the zero-section and then $\{e_4=0\}$ is an order two generator of the Mordell-Weil group.
It is a consequence of the torsional section that there are no  hypermultiplets in the fundamental representation that would arise from isolated fibral curves over $C_{G}$~\cite{Aspinwall:1998xj,Mayrhofer:2014opa}.

The fact that $\{e_4=0\}$ and $\{e_5=0\}$ are holomorphic sections implies also that
\begin{align}
	E_1^2=-c_1(B)E_1\,,\quad E_2^2=-c_1(B)E_2\,,
\end{align}
and from \eqref{eqn:p1}, $s_9=1$ and the Stanley-Reisner ideal of $\mathbb{P}_{F_{13}}$ it follows that $E_1\cdot E_2=0$.
We denote the intersections on $B$ by
\begin{align}
	c_{ij}=\int\limits_B D_i\cdot D_j\,.
\end{align}
Together this determines the intersections on $M$.
For later convenience let us denote the curve on $B$ that is dual to $\tilde{D}_i$ by $C^i$, i.e.
\begin{align}
	\tilde{D}_i\cdot C^j=\delta_i^j\,.
\end{align}
for $i,j=1,...,h^{1,1}(B)$ and introduce
\begin{align}
	a=\int\limits_B c_1(B)^2\,,\quad a^i=\int\limits_Bc_1(B)\cdot C^i\,,\quad a_i=\int\limits_Bc_1(B)\cdot \tilde{D}_i.
\end{align}

We will now construct a set of branes in $D^b(M)$ such that the central charges generate the charge lattice.
As the six-brane we use the structure sheaf $\mathcal{O}_M$ on $M$.
From the Shioda-Tate-Wazir theorem~\cite{TateWazir} we know that $H^{1,1}(M)$ is generated by $E_1,E_2$ and the vertical divisors $D_i$ with
\begin{align}
	D_i=\pi^*\tilde{D}_i\quad\text{for}\quad i=1,...,h^{1,1}(B)\,,
\end{align}
where the $\tilde{D}_i$ form a basis of $H^{1,1}(B)$.
To construct four-branes we then use the standard short exact sequence for Cartier divisors which implies
\begin{align}
	\mathcal{O}_D\sim\mathcal{O}_M(-D)\rightarrow\mathcal{O}_M\,,
\end{align}
as elements of the derived category and take the free resolutions of $\mathcal{O}_D$ for $D\in\{E_1,E_2,D_i,\,i=1,...,h^{1,1}(B)\}$.
As a basis of 2-branes on $M$ we consider the curves $C_F=C_A+C_B,C_B$ and $C_i=E_1\cdot D_i,\,i=1,...,h^{1,1}(B)$ and the corresponding K-theoretic push-forwards~\cite{Gerhardus:2016iot}
\begin{align}
	\mathcal{C}^\bullet=\iota_{!}\mathcal{O}_{C}(K_C^{1/2})\,.
\end{align}
Zero-branes are described by skyscraper sheaves $\mathcal{O}_{p}$ with support on some point ${p}\in M$.
The corresponding Chern characters are
\begin{align}
	\begin{split}
	\text{ch}(\mathcal{O}_M)=&1\,, \quad \text{ch}(\mathcal{O}_D)=1-\exp(-D)\,,\\ 
		\text{ch}(\mathcal{C}^\bullet)=&C\,,\quad \text{ch}(\mathcal{O}_{p})=V\,,
	\end{split}
\end{align}
where $V$ is the fundamental class of $M$.

To calculate the action of the relative Fourier-Mukai transforms we follow~\cite{Andreas:2004uf} and generalize the calculation to geometries with reducible fibers.
To this end it will be crucial that we constructed $M$ in an ambient space $\widehat{\mathbb{P}}_{F_{13}}$ that consists of a toric variety fibered over the base $B$
and therefore $i:M\hookrightarrow \widehat{\mathbb{P}}_{F_{13}}$ is a local complete intersection (l.c.i.) morphism in the sense of \cite{fulton1984intersection}.
This implies that the action of a relative Fourier-Mukai transform $\Phi_{\mathcal{E}}$ over $B$ with kernel $\mathcal{E}$ on the Chern character of a brane $\mathcal{F}^\bullet$ is given by
\begin{align}
	\text{ch}\left(\Phi(\mathcal{F}^\bullet)\right)=\pi_{1,*}\left[\text{ch}(\mathcal{E})\pi^*_2\left(\text{ch}(\mathcal{F}^\bullet)\cdot\text{Td}\left(T_{M/B}\right)\right)\right]\,.
\end{align}
Here the projections correspond to the diagram
\begin{equation}
\xymatrix{
	M\times_B M\ar[d]^{\pi_1}\ar[r]^{\pi_2}\ar[rd]^{p}&M\ar[d]_{\pi}\\
	M \ar[r]^{\pi}	&B
	}\,,
\label{eqn:fmdiagram}
\end{equation}
and $T_{M/B}$ is the so-called \textit{virtual relative tangent bundle} of $\pi:M\rightarrow B$.

The latter is not an actual bundle but one can associate to it the K-class
\begin{align}
	[T_{M/B}]=[i^*T_{P/B}]-[N_{M/P}]\,.
\end{align}
We calculate the corresponding Chern class
\begin{align}
	\text{c}(T_{M/B})=\frac{(1+E_1)(1+E_2)(1+2E_1-2E_2+2c_1(B)(1+E_1+E_2+c_1(B))}{1+4E_1+4c_1(B)}\,,
\end{align}
and from this
\begin{align}
	\text{Td}\left(T_{M/B}\right)=1-\frac{1}{2}c_1(B)+\frac{1}{12}c_1(B)(7c_1(B)+6E_1+6E_2)-\frac{1}{2}c_1(B)^2E_1\,.
\end{align}
Note that the leading behaviour $\text{Td}\left(T_{M/B}\right)=1-\frac12c_1(B)+\dots$ is an inherent consequence of the construction and independent of the particular class of fibers.

We now specialize to the case where $\mathcal{E}=\mathcal{I}_\Delta$ is the ideal sheaf of the relative diagonal $\delta:M\hookrightarrow M\times_B M$.
Let us first note that~\cite{Andreas:2004uf}
\begin{align}
	\text{ch}(\mathcal{I}_\Delta)=1-\text{ch}(\delta_*\mathcal{O}_M)
\end{align}
and use singular Riemann-Roch~\cite{fulton1984intersection,Andreas:2004uf} to obtain
\begin{align}
	\text{ch}(\delta_*\mathcal{O}_M)\text{Td}(M\times_B M)=\delta_*(\text{ch}(\mathcal{O}_M)\text{Td}(M))\,.
\end{align}
Then we get the result
\begin{align}
	\text{ch}\left(\Phi(\mathcal{F}^\bullet)\right)=\text{ch}(\mathcal{F}^\bullet)-\pi_{2,*}\left[\pi_1^*\left(\text{ch}(\mathcal{F}^\bullet)\text{Td}(T_{M/B})\right)\right]\,.
	\label{eqn:result1}
\end{align}
The second term in \eqref{eqn:result1} can be evaluated by noting that the push-forward operation in cohomology is adjoint to the pull-back.
It is easisest to describe the corresponding action on the dual cycles.
The pull-back along $\pi_1^*$ transforms the dual cycle of a form into a cycle that wraps the ``left'' fiber of $M\times_B M$ and $\pi_{2,*}$ projects out forms where the dual cycle is not pointlike on the ``right'' fiber.
In other words the action is
\begin{align}
	\pi_{2,*}\pi_1^*:\,\left\{\begin{array}{l}1\mapsto 0\,,\\
		E_1\mapsto 1,\,\quad E_2\mapsto 1\,,\quad D_i\mapsto 0,\,i=1,...,h^{1,1}(B)\,,\\
		C_F\mapsto 0\,,\quad C_B\mapsto 0\,,\quad C_i\mapsto D_i,\,i=1,...,h^{1,1}(B)\,,\\
		V\mapsto C_F
	\end{array}\right.\,.
\end{align}

With this we can finally write down the action of $\Phi_{\mathcal{I}_\Delta}$ on our basis of branes.
To this end we arrange the charges into a vector
\begin{align}
	\begin{split}
	\vec{\Pi}=&\left(\text{ch}(\mathcal{O}_M),\,\text{ch}(\mathcal{O}_{E_1}),\,\text{ch}(\mathcal{O}_{E_2}),\,\text{ch}(\mathcal{O}_{D_1}),\,\dots,\,\text{ch}\left(\mathcal{O}_{D_{h^{1,1}(B)}}\right),\right.\\
		&\left.\text{ch}(\mathcal{C}^\bullet_1),\,\dots,\,\text{ch}\left(\mathcal{C}^\bullet_{h^{1,1}(B)}\right),\,\text{ch}\left(\mathcal{C}^\bullet_F\right),\,\text{ch}\left(\mathcal{C}^\bullet_B\right),\,\text{ch}(\mathcal{O}_p)\right)\,,
	\end{split}
\end{align}
and find that $\Phi_{\mathcal{I}_\Delta}:\vec{\Pi}\mapsto S\cdot\vec{\Pi}$, with
\begin{align}
S=\begin{blockarray}{rrrrccccl}
	\mathcal{O}_M&\mathcal{O}_{E_1}&\mathcal{O}_{E_2}&\mathcal{O}_{D_i}&\mathcal{C}_i^\bullet&\mathcal{C}^\bullet_F&\mathcal{C}^\bullet_B&\mathcal{O}_p&\\
\begin{block}{(rrrrcccc)l}
	1&0&0&-a^i&0&\tilde{a}&0&0&\mathcal{O}_M\\
		-1&1&0&0&0&0&0&0&\mathcal{O}_{E_1}\\
		-1&0&1&0&0&0&0&0&\mathcal{O}_{E_2}\\
		0&0&0&\delta^i_j&0&-a_j&0&0&\mathcal{O}_{D_j}\\
		0&0&0&-1&1&\frac{1}{2}(a_j-c_{jj})&0&0&\mathcal{C}^\bullet_j\\
		0&0&0&0&0&1&0&0&\mathcal{C}^\bullet_F\\
		0&0&0&0&0&0&1&0&\mathcal{C}^\bullet_B\\
		0&0&0&0&0&-1&0&1&\mathcal{O}_{p}\\
\end{block}
\end{blockarray}\,.
\label{eqn:f13transform}
\end{align}
where we introduced $\tilde{a}=\frac12\left(a-a^mc_{mm}\right)$.

\subsection{$G=SU(2)\,,\quad \#n_F=8c_1(B)^2$}
As a second example we will now calculate the transformation for Calabi-Yau threefolds with a curve of $SU(2)$ singularities and with hypermultiplets in the fundamental representation.
The toric ambient space of the fiber will be $\mathbb{P}_{F_{10}}$  with the data
\begin{align}
\begin{blockarray}{crrrrl}
	&&&C_F^1&C_F^2&\\
\begin{block}{r(rr|rr)l}
	e_3& -1&-2&-1& 1&\leftarrow\text{ holomorphic section }E_1\\
	v& -1& 1& 1& 0&\leftarrow\text{ two-section }E_2\\
	w&  1& 0& 0& 1&\leftarrow\text{ three-section}\\
	u&  0&-1& 3&-2&\leftarrow\text{ fibral divisor }D_f\\
	&  0& 0&-3& 0&\\
\end{block}
\end{blockarray}\,.
	\label{eqn:toricdataf13}
\end{align}
and we promote this to a fibration over a surface $B$ by letting the homogeneous coordinates be sections of line bundles on $B$.
The Stanley-Reisner ideal is
\begin{align}
	\mathcal{SRI}=\langle v u,\,w e_3\rangle.
\end{align}
Following \cite{Klevers:2014bqa} we parametrize the classes of the homogeneous coordinates again by $\mathcal{S}_7,\mathcal{S}_9\in H^{1,1}(B)$ such that
\begin{align}
	\begin{split}
		[u]=&E_2-2E_1+\mathcal{S}_7-c_1(B)\,,\quad[v]=E_2\,,\\
		[w]=&E_1+E_2-\mathcal{S}_9+\mathcal{S}_7\,,\quad [e_3]=E_1\,.
	\end{split}
\end{align}
The generic section of the anti-canonical bundle on the total space $\widehat{\mathbb{P}}_{F_{10}}$ reads
\begin{align}
	p_{F_{10}}=s_1e_3^6u^3+s_2e_3^4u^2v+s_3e_3^2uv^2+s_4v^3+s_5e_3^3u^2w+s_6e_3uvw+s_8uw^2\,,
\end{align}
and the Calabi-Yau condition forces~\cite{Klevers:2014bqa}
\begin{align}
	\begin{split}
	[s_1]=&3c_1(B)-\mathcal{S}_7-\mathcal{S}_9\,,\quad [s_2]=2c_1(B)-\mathcal{S}_9\,,\quad [s_3]=c_1(B)+\mathcal{S}_7-\mathcal{S}_9\,,\\
		[s_4]=&2\mathcal{S}_7-\mathcal{S}_9\,,\quad[s_5]=2c_1(B)-\mathcal{S}_7\,,\quad[s_6]=c_1(B)\,,\quad [s_8]=c_1(B)+\mathcal{S}_9-\mathcal{S}_7\,.
	\end{split}
\end{align}

For $\mathcal{S}_7=c_1(B),\,\mathcal{S}_9=0$ the ambient space does not have to be resolved any further and we obtain a Calabi-Yau $M$ with the desired properties.
In particular there is a genus $g=1+c_1(B)^2$ curve $C_G$ of $I_2$ fibers that are resolved by the fibral divisor $D_f$.
Again we will denote the components of the generic fiber over $C_G$ by $C_A$ and $C_B$ such that $C_A$ is transversely intersected by $E_1$.
The two-section $\{v=0\}$ with $[v]=E_2$ will intersect each of the components $C_A$ and $C_B$ once.
Over $8c_1(B)^2$ points on $C_G$ the fiber degenerates further into an $I_3$ configuration.
In an F-theory compactification this leads to a corresponding number of hypermultiplets in the fundamental representation of $SU(2)$.

We will denote the generic fiber of $M$ by $C_F=C_A+C_B$.
As a basis for $H^{1,1}(M)$ we choose $E_1,E_2$ and the vertical divisors $D_i=\pi^*\tilde{D}_i,\,i=1,...,h^{1,1}(B)$.
The intersections are determined by
\begin{align}
	E_1^2=-c_1(B)E_1\,,\quad E_1E_2=E_2^2=0\,,
\end{align}
and by the intersections on the base.
For the latter we use the notations that were introduced in the first example.
The Chern class of the virtual relative tangent bundle is now
\begin{align}
	\text{c}(T_{M/B})=\frac{(1+E_1)(1+E_2)(1+E_1+E_2+c_1(B))(1+E_2-2E_1)}{1+3E_2+2c_1(B)}\,,
\end{align}
and we calculate the corresponding Todd class
\begin{align}
	\text{Td}\left(T_{M/B}\right)=1-\frac12c_1(B)+\frac{1}{12}c_1(B)(3c_1(B)+2E_1+5E_2)-\frac12c_1(B)^2E_1\,.
\end{align}
We choose our basis of branes
\begin{align}
	\begin{split}
	\vec{\Pi}=&\left(\text{ch}(\mathcal{O}_M),\,\text{ch}(\mathcal{O}_{E_1}),\,\text{ch}(\mathcal{O}_{E_2}),\,\text{ch}(\mathcal{O}_{D_1}),\,\dots,\,\text{ch}\left(\mathcal{O}_{D_{h^{1,1}(B)}}\right),\right.\\
		&\left.\text{ch}(\mathcal{C}^\bullet_1),\,\dots,\,\text{ch}\left(\mathcal{C}^\bullet_{h^{1,1}(B)}\right),\,\text{ch}\left(\mathcal{C}^\bullet_F\right),\,\text{ch}\left(\mathcal{C}^\bullet_B\right),\,\text{ch}(\mathcal{O}_p)\right)\,,
	\end{split}
\end{align}
and find that $\Phi_{\mathcal{I}_\Delta}:\vec{\Pi}\mapsto S\cdot\vec{\Pi}$, with
\begin{align}
S=\begin{blockarray}{rrrrccccl}
	\mathcal{O}_M&\mathcal{O}_{E_1}&\mathcal{O}_{E_2}&\mathcal{O}_{D_i}&\mathcal{C}_i^\bullet&\mathcal{C}^\bullet_F&\mathcal{C}^\bullet_B&\mathcal{O}_p&\\
\begin{block}{(rrrrcccc)l}
	1&0&0&-a^i&0&\tilde{a}&0&0&\mathcal{O}_M\\
		-1&1&0&0&0&0&0&0&\mathcal{O}_{E_1}\\
		-2&0&1&a^i&0&-\tilde{a}&0&0&\mathcal{O}_{E_2}\\
		0&0&0&\delta^i_j&0&-a_j&0&0&\mathcal{O}_{D_j}\\
		0&0&0&-1&1&\frac{1}{2}(a_j-c_{jj})&0&0&\mathcal{C}^\bullet_j\\
		0&0&0&0&0&1&0&0&\mathcal{C}^\bullet_F\\
		0&0&0&0&0&0&1&0&\mathcal{C}^\bullet_B\\
		0&0&0&0&0&-1&0&1&\mathcal{O}_{p}\\
\end{block}
\end{blockarray}\,.
\label{eqn:f13transform}
\end{align}
where we introduced $\tilde{a}=\frac12\left(a-a^mc_{mm}\right)$.

\section{Modularity from monodromy}
\label{sec:3}
We now want to show that the relative conifold transformations lead to an action of the modular group on the flat coordinates.
In this section we will refer to a general elliptic Calabi-Yau threefold that is compatible with our construction by $M$.
We stress that this discussion is \textit{not} restricted to the examples that we discussed in the previous section.

From the definition of the virtual relative  tangent bundle it is clear that the leading terms of $\text{Td}(T_{M/B})$ will always be
\begin{align}
	\text{Td}(T_{M/B})=1-\frac{1}{2}c_1(B)+\dots\,,
	\label{eqn:leadingtodd}
\end{align}
up to terms of degree $2$ or higher.
This implies that the central charges of 2-branes wrapping fibral curves will always be unaffected by this transformation.
On the other hand 2-branes that wrap curves in the base will transform into a bound-state of the 2-brane itself, a 4-brane wrapping the corresponding vertical divisor and
a certain number of 2-branes that wrap the generic fiber of the fibration.
More precisely one finds that
\begin{align}
	\Phi:\,\text{ch}(\mathcal{C}_i^\bullet)\mapsto\text{ch}(\mathcal{C}_i^\bullet)-\text{ch}(\mathcal{O}_{D_i})+\frac{1}{2}(a_i-c_{ii})\text{ch}(\mathcal{C}_F^\bullet)\,.
	\label{eqn:basecurvetrans}
\end{align}
We stress that this is not restricted to our particular examples but follows in general from \eqref{eqn:leadingtodd}.
It will also be true in general that
\begin{align}
	\Phi:\,\text{ch}(\mathcal{O}_{p})\mapsto\text{ch}(\mathcal{O}_{p})-\text{ch}(\mathcal{C}_F^\bullet)\,.
\end{align}

The asymptotic behaviour of the central charge of a B-brane $\mathcal{F}^\bullet$ in the limit of large volume can be calculated using the $\Gamma$-class formula~\cite{Iritani:2009}
\begin{align}
	\Pi_{\text{asy}}(\mathcal{F}^\bullet)=\int\limits_M e^{\omega}\Gamma_{\mathbb{C}}(M)(\text{ch}\,\mathcal{F}^\bullet)^\vee\,,
	\label{eqn:bbranecharge}
\end{align}
where in terms of the Chern classes $c_2,c_3$ of $M$
\begin{align}
	\Gamma_{\mathbb{C}}(M)=1+\frac{1}{24}c_2+\frac{\zeta(3)}{(2\pi i)^3}c_3\,.
\end{align}
The action of the operator $\vee:\oplus_k H^{k,k}(M)\rightarrow\oplus_k H^{k,k}(M)$ is linear and determined by $\delta^\vee\mapsto(-1)^i\delta$ for $\delta\in H^{i,i}(M)$.
We also introduced the complexified K\"ahler form $\omega$ that we can expand as
\begin{align}
	\omega=\sum\limits_{i=1}^{h^{1,1}(M)}t^iD_i\,,
\end{align}
in terms of divisors $D_i\in H^{1,1}(M)$.

We will now assume that the fibration admits a section.
It follows from the Shioda-Tate-Wazir theorem~\cite{TateWazir} that $H^{1,1}(M)$ is generated by the zero section $E_0$, the divisors that correspond to free generators of the Mordell-Weil group $E_i,\,i=1,...,k=\text{rk}(MW)$,
exceptional divisors $D^f_i,\,i=1,...,g$ that resolve curves of singular fibers and vertical divisors $D_i=\pi^*\tilde{D}_i,\,i=1,...,h^{1,1}(B)$.
We also introduce $\tilde{E}_0=E_0+c_1(B)$ and the images under the Shioda map
\begin{align}
	¸\tilde{E}_i=\sigma(E_i)=E_i-{E}_0-c_1(B)-\pi^{-1}\pi(E_i\cdot {E}_0)+\tilde{D}^f\,,
\end{align}
where $\tilde{D}^f$ is a linear combination of fibral divisors such that $\tilde{E}_i\cdot D^f_j=0$ for all $i=1,...,k$ and $j=1,...,g$.
The Shioda map is constructed such that the image is orthogonal to the subspace generated by $\tilde{E}_0$, the fibral divisors $D^f_i$ and the vertical divisors $D_i$~\cite{Morrison:2012ei}.
We can now expand the K\"ahler form as
\begin{align}
	\omega=\tau \tilde{E}_0+\sum_{i=1}^km^i\tilde{E}_i+\sum\limits_{i=1}^gn^iD^f_i +\sum\limits_{i=1}^{h^{1,1}(B)}\tilde{t}^iD'_i,.
\end{align}
such that the vertical divisors $D_i'$  are dual to the curves $C_i=E_0\cdot D_i$.
Then $\tau$ is the complexified volume of the generic fiber, $m^i$ are the volumes of isolated fibral curves and $n^i$ are linear combinations of the volumes of fibers of fibral divisors.

From the previous discussion it follows that the complexified volume $m$ of a fibral curve transforms as
\begin{align}
	\Phi:\,m\mapsto\frac{m}{1+\tau}\,.
\end{align}
On the other hand, for a 2-brane that wraps a curve $C_i=E_0\cdot D_i$ we use \eqref{eqn:basecurvetrans} and find that the central charge transforms as
\begin{align}
	\begin{split}
		\Phi:\,\tilde{t}^i\mapsto&\frac{1}{1+\tau}\left((1+\tau)\tilde{t}^i+\frac12a_i\tau^2+\frac12\tau c_{ii}+\frac{1}{24}\int\limits_M c_2(M)D_i+\frac12(a_i-c_{ii})\tau\right.\\
		&\left.+\frac12m^am^b\int\tilde{E}_a\tilde{E}_bD_i'+\frac12 n^an^b\int D^f_aD^f_bD_i'\right)\\
		=&\tilde{t}^i+\frac12 a_i\tau +\frac{1}{1+\tau}\frac{1}{24}\int\limits_Mc_2(M)D_i\\
		&+\frac12\frac{1}{1+\tau}\left(m^am^b\int\limits_M \tilde{E}_a\tilde{E}_bD_i+n^an^b\int\limits_M D^f_aD^f_bD_i\right)\,,
	\end{split}
\end{align}
up to quantum corrections that are exponentially surpressed in the limit of large base~\footnote{This follows from the fact that the central charge of the 4-brane that is substracted in \eqref{eqn:basecurvetrans} is the partial derivative
of a prepotential with respect to the complexified volume of a curve in the base.}.

To manifestly see the action of the modular group we twist the branes with the line bundle $\mathcal{O}(-\tilde{E}_0)$ before and after the transformation.
Note that such a twist is generated by the corresponding large volume monodromies and manifestly an autoequivalence of the category of B-branes.
We denote the resulting transformation by $S$ and observe
\begin{align}
	S:\quad \tau\mapsto -\frac{1}{\tau}\,,\quad \tilde{t}^i\mapsto \tilde{t}^i+\frac12 a_i(\tau-1)+\frac{1}{\tau}\frac{\tilde{c}_i}{24}-\frac12\frac{1}{\tau}\left(m^am^bC_{ab}^i+n^an^b\tilde{C}^i_{ab}\right)\,,
\end{align}
where we introduced
\begin{align}
	\tilde{c}_i=\int\limits_Mc_2(M)D_i\,,\quad C_{ab}^i=-\int\limits_M\tilde{E}_a\cdot\tilde{E}_c\cdot D_i\,,\quad\tilde{C}_{ab}^i=-\int\limits_M D^f_aD^f_bD_i\,.
\end{align}
Here $C_{ab}^i$ is the so-called height pairing of sections $E_a$ and $E_b$.
The complexified volume $m$ of any fibral curve transforms as
\begin{align}
	S:\,m\mapsto\frac{m}{\tau}\,.
\end{align}

In the examples that we presented one can easily calculate that
\begin{align}
	\tilde{c}_i=\int\limits_Mc_2(M)D_i=\int\limits_Bc_1(B)\tilde{D}_i=a_i\,,
\end{align}
for any choice of base $B$. We do not know how to derive this relation in general but let us assume that it holds.
We can then introduce the shifted K\"ahler moduli
\begin{align}
	t^i=\tilde{t}^i+\frac{a_i}{2}\tau\,,
\end{align}
as well as the exponentials $Q^i=\exp(2\pi i t^i)$ and find the result
\begin{align}
	S:\,\left\{\begin{array}{rl}
		\tau&\mapsto-\frac{1}{\tau}\\
		m&\mapsto\frac{m}{\tau}\\
		Q^i&\mapsto(-1)^{a_i}\exp\left(-\frac{1}{\tau}\left(m^am^bC_{ab}^i+n^an^b\tilde{C}^i_{ab}\right)\right)Q^i
	\end{array}\right.\,.
	\label{eqn:modtrans1}
\end{align}
Here $m$ stands again  for the complexified volume of any fibral curve.
We see that the volume of the generic fiber transforms like a modular parameter,
the complexified volumes of rational fibral curves transform like elliptic parameters and the exponentiated volumes of base curves transform like lattice Jacobi forms of weight $0$ and with index matrices $C^i_{ab}$ and $\tilde{C}^i_{ab}$.
\section{Conclusion}
In this note we have explicitly calculated the action of a relative conifold transformation on a basis of central charges of B-branes for two classes of elliptically fibered 
Calabi-Yau varieties.
In contrast to previous calculations in the literature our examples exhibit reducible fibers.
Moreover, an analogous calculation can be performed for any elliptically fibered Calabi-Yau where the fiber is a complete intersection in a toric ambient space.
In particular we do not require the existence of a section.

From the explicit calculations we then deduced the action on the 2-brane charges of arbitrary elliptically fibered Calabi-Yau threefolds of this type.
The result depends on whether the fibration admits a section or only $k$-sections.
For geometries with a section we found an action of the modular group.
The volume of the fiber transforms like a modular parameter while the volumes of rational fibral curves transform like elliptic parameters of Jacobi forms.
The exponential of appropriately shifted base parameters transforms up to a multiplier system like a lattice Jacobi form of weight $0$.
The index matrix is block diagonal such that the volumes of isolated fibral curves and fibers of fibral divisors do not mix.
The indices of the former correspond to the height pairing between the sections while the indices of the latter are essentially given by the intersection of the corresponding fibral divisors.

In a companion paper that is currently being finalized we will, together with other authors, connect this with the modular anomaly equations and to the modular properties of the topological string partition function~\cite{wip}.
There we will also discuss the action on the flat coordinates for elliptic fibrations that do not admit a section.

Apart from the relation to modular properties of the topological string partition function we expect that our results could also be useful in the context of the spectral cover construction~\cite{Friedman:1997yq}.
\section*{Acknowledgments} We want to thank Albrecht Klemm, Johanna Knapp and Cesar Fierro Cota for helpful discussions.
The work of the author is supported by the Austrian Science Fund (FWF): P30904-N27.
\addcontentsline{toc}{section}{References}
\bibliographystyle{utphys}
\bibliography{names}
\end{document}